\documentclass[prl,aps,twocolumn,epsfig,10pt]{revtex4-2}
\usepackage{graphicx}% Include figure files
\usepackage{dcolumn}% Align table columns on decimal point
\usepackage{bm}% bold math
\usepackage{lipsum}
\usepackage{xcolor}
\usepackage{soul}
\usepackage{ulem}
\usepackage[utf8]{inputenc}
\usepackage{newunicodechar}
\newunicodechar{₂}{$_2$}
\makeatletter

\newcommand{\Rmnum}[1]{\expandafter\@slowromancap\romannumeral #1@}
\makeatother
\begin{document}

\title {
Polarization engineered all 2D Graphene/Ferroelectric hybrid for persistence-free photoresponse 
}
\author{Navkiranjot Kaur Gill$^{1*}$, Shaili Sett$^{2}$, Saloni Kakkar$^{1}$, Kenji Watanabe$^{3}$, Takashi Taniguchi$^{4}$ and Arindam Ghosh$^{1,5}$}
\vspace{1.5cm}
\affiliation{$^1$Department of Physics, Indian Institute of Science, Bangalore 560012, India.}
\affiliation{$^2$ Centre for Interdisciplinary and Convergent Technologies, Indian Institute of Technology, Kharagpur 721302, India.}
\affiliation{$^3$Research Center for Electronic and Optical Materials, National Institute for Materials Science, 1-1 Namiki, Tsukuba 305-0044, Japan.}
\affiliation{$^4$Research Center for Materials Nanoarchitectonics, National Institute for Materials Science,  1-1 Namiki, Tsukuba 305-0044, Japan.}
\affiliation{$^5$Centre for Nanoscience and Engineering, Indian Institute of Science, Bangalore 560012, India.} 
\affiliation{$^*$email-id: navkiranjotg@iisc.ac.in}

\begin{abstract}


Graphene–transition metal dichalcogenide hybrid photodetectors typically work on trap-mediated photogating mechanism, exhibiting high sensitivity, but under-perform in the fast detection of repetitive optical signals. Designing photodetectors that are simultaneously fast and highly sensitive has therefore remained difficult. In this work, we realize both attributes by integrating atomically thin sliding ferroelectrics in the design architecture, thereby uniting semiconducting properties with intrinsic polarization fields capable of efficiently governing interfacial photocarrier dynamics. We report a bilayer graphene/bilayer MoS$_2$ (with MoS$_2$ in a rhombohedrally stacked (3R) configuration) van der Waals photodetector with edge-contacted dual-gated field-effect transistor architecture. The photo-induced modulation in spontaneous out-of-plane polarization of 3R-MoS$_2$, and selective confinement of charge carriers in bilayer graphene under an out-of-plane displacement field results in a tunable persistence-free photoresponse. Here, the photoinduced polarization change in 3R-MoS$_2$ produces an optically controlled gating effect that alters the electrostatic environment of bilayer graphene, resulting in a temperature-independent photoresponse with rapid response times of the order of 10's of milliseconds (limited by the measurement instrument). This polarization-assisted photoresponse enables measurement of repeatable optical signal detection in high gain regime. We demonstrate reproducible detection of optical signals and examine the photon-counting resolution of this structure in high-sensitivity regimes, where we determine its internal quantum efficiency to be $\approx$ $10\%$, with minimum detectable photon number of 31 in single shot measurements. This work highlights the functionality of 3R-MoS$_2$ in manipulating the interfacial charge dynamics and establishes the hybrid of graphene and ferroelectric 3R-MoS$_2$ as a promising platform for ultra-sensitive optoelectronic devices.

\end{abstract}

\maketitle

\section{I. Introduction}
Transition metal dichalcogenides (TMDs) have attracted considerable attention as promising photodetector materials since their initial discovery, owing to exceptionally high optical absorption coefficients that are retained even at the atomically thin limit \cite{bernardi2013extraordinary}. Over the past decade, two-dimensional (2D) material-based photodetectors has undergone significant transformations through the realization of van der Waals heterostructures formed by vertically stacking different 2D layers. Such hybrids like  Graphene (Gr)/TMD have enabled ultrahigh photoresponsivities of $5 \times 10^{10}$ AW$^{-1}$ \cite{roy2013graphene}, achieved through trap-mediated photogating effect. The exceptionally large optical gain originates from two cooperative physical mechanisms: first, the energetically favorable band alignment at the Gr/TMD interface which drives efficient photoinduced charge transfer, and second, the substantially prolonged carrier lifetime arising from trapping of photoexcited carriers in localized states associated with traps. These include interfacial trap states, sulfur vacancy defects intrinsic to the TMD lattice, and charge traps introduced by molecular adsorbates on 2D surfaces. Due to the large de-trapping timescales, Gr/TMD hybrid photodetectors exhibit persistent photoconductivity extending over timescales ranging from minutes to days at low temperatures ($T<$150 K) \cite{ahmed2020interplay,gill2025moire,parappurath2022interlayer,kashid2020observation,roy2013graphene,mitra2020graphene} imposing severe constraints on detector response speed and necessitating the application of a large reset gate voltage pulse to restore the detector to its equilibrium dark-state conductance \cite{roy2013graphene}.

\begin{figure*} 
\includegraphics[width=1\linewidth]{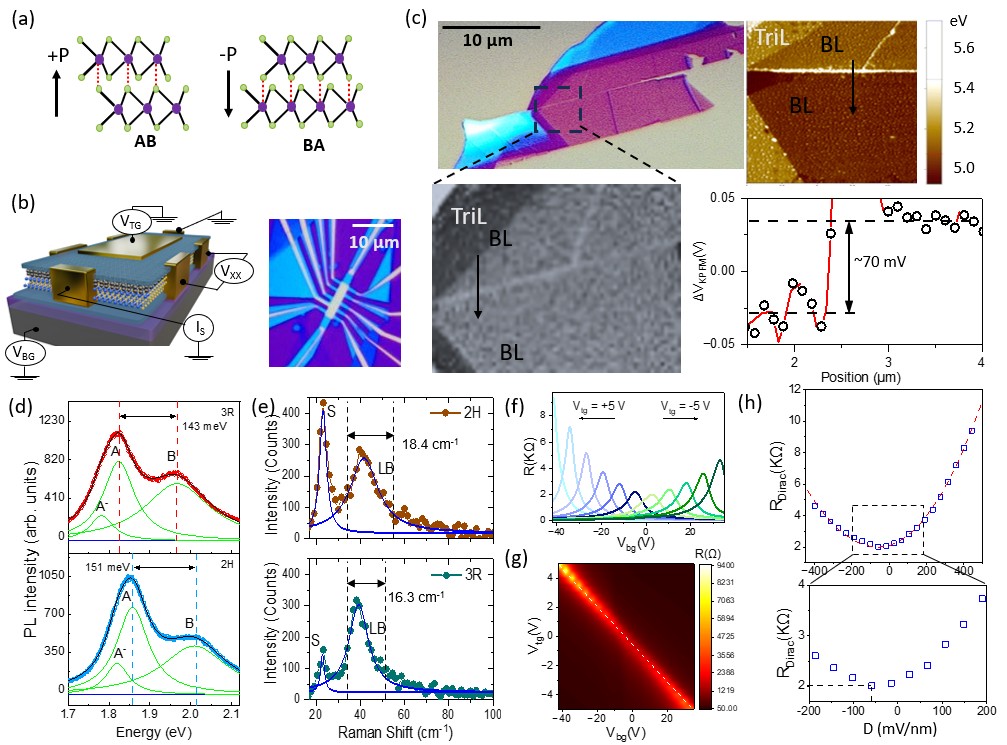} 
\caption{Characterization of BLG/3R-MoS$_2$ FET. (a) Two different stacking configuration of 3R polytype, $AB$ and $BA$, with arrows depicting the direction of spontaneous out-of-plane polarization ($P$). (b) Schematic illustration of the device architecture, comprising a hexagonal boron nitride (hBN)–encapsulated bilayer graphene (BLG) on a 3R-MoS$_{2}$, equipped with a gold top gate and a hBN/SiO$_2$ back gate, configured in a hall-bar geometry with one-dimensional edge contacts to the BLG. The right panel shows an optical micrograph of the fabricated device. (c) KPFM measurement microgarphs: optical image of a 3R MoS$_2$ multilayer flake. The false color zoomed region shows trilayer (TriL) and bilayer (BL) portions of the flake, while the white contrast region is a crack in the flake. The surface potential map alongside shows two distinct regions in the bilayer portion. The arrow indicates the line cut map taken as shown below. $\Delta V_{KPFM}$ is plotted as a function of position indicated by the arrow above. (d) PL spectra of 3R and 2H- MoS$_2$ at room temperature showing multiple excitonic peaks. The dashed lines indicate the shift in peak positions of $A$ and $B$ exciton for these two stacking configurations. (e) Low wavenumber Raman spectra of 2H- and 3R- MoS$_2$ showing prominent layer breathing ($LB$) and shear modes ($S$). The frequency difference between the modes is indicated by dashed lines. (f) $R-V_{bg}$ at 100 K in BLG/3R-MoS$_2$ at different $V_{tg}$ from -5 V to +5 V in steps of 1 V. (g) The color plot of $R$ versus $V_{bg}$-$V_{tg}$. (h) Dirac point resistance ($R_{Dirac}$) as a function of displacement field ($D$), which a zoomed region near the minima of $R_{Dirac}$.} 
\label{1}
\end{figure*}

Here we propose a strategy based on the incorporation of the rhombohedral polytype (3R) of TMDs that exhibit sliding ferroelectricity \cite{li2017binary, sett2025van}. 3R-TMDs represent an emerging class of layered ferroelectrics, with an inversion -symmetry broken non-centrosymmetric stacking. The dipole formation at the interface of two consecutive TMD layers leads to spontaneous out-of-plane polarization that persists till room temperature \cite{wang2022interfacial,weston2022interfacial,sett2024emergent}. The two such stacking configurations are- 1) $AB$: where the metal atom of the top layer sits on the chalcogen atom of bottom layer and 2) $BA$: where the chalcogen atom of top layer sits on the metal atom of bottom layer (see Fig.\ref{1}a). These can be distinguished by a relative lateral interlayer shift of one-third of the lattice parameter and can be switched into one another by applying an out-of-plane external field \cite{ko2023operando}. Other notable features of the 3R-stacking are: (1) asymmetric interlayer coupling that induces layer-resolved polarization of the electronic wavefunction and a staggered band alignment at the $K$ point of the Brillouin zone \cite{wang2017interlayer,yang2022spontaneous,liang2022optically}; (2) enhancement of excitonic lifetime attributable to the spatial separation of charge carriers across adjacent layers \cite{scuri2020electrically,choi2021twist}; and (3) a statistically higher probability of interfacial sulfur vacancy formation \cite{wan2026defect}.

Prior investigations on the 3R-MoS$_2$ have demonstrated an enhanced external quantum efficiency of up to 16$\%$ in vertical photovoltaic device architectures, where MoS$_2$ is sandwiched between two graphene electrodes, indicating an efficient dissociation of excitons in 3R-MoS$_2$ compared to the natural bilayer 2H-stacking \cite{yang2022spontaneous,wu2022ultrafast}. Building upon these findings, the present work examines bilayer graphene (BLG)/3R-MoS$_2$ (bilayer) encapsulated with hexagonal boron nitride (hBN) configured in a field-effect transistor architecture. Here, BLG serves as the charge transport channel, enabling a systematic investigation of the influence of $\mu$m-scale ferroelectric polar domains on the kinetics of interfacial charge transfer across the Gr/TMD hetero-interface. A key finding of this work is a persistence-free photoresponse at low temperatures ($T <$ 150 K), exhibiting recovery timescale of about 22 ms (limited by measurement instrument), suggesting a qualitative change in the underlying charge transfer dynamics. The Gr/3R-MoS$_2$ hybrid show photon-number-resolving capability where absorbing 262 photons by MoS$_2$ give 31 photon counts, demonstrating ultra-low light detection. In addition, we obtain a Gain-Bandwidth (GBW) product of $\approx 10^{8}$ Hz, one of the highest among photodetectors built on the principle of trap-mediated photogating effect.

\section{II Experimental Details}
2D layers of graphene, 3R-MoS$_2$ and hBN were mechanically exfoliated using the scotch-tape technique on SiO$_2$$\slash$Si$^{++}$ substrates \cite{novoselov2004electric}. Bilayers of graphene and MoS$_2$ were identified using Raman and Photolumniscense (PL) Spectroscopy. 3R-MoS$_2$ bilayer flakes are characterized using Kelvin Probe force microscopy (KPFM) in frequency modulation mode. KPFM mapping is done at an amplitude of 11 nm and a AC bias of 2 V at 1.5 KHz frequency. For fabricating the heterostructure, we use polycarbonate (PC) film as a sacrificial layer on a PDMS hemispherical drop to pick-up and release the 2D flakes by dry-transfer technique \cite{wang2013one}. The heterostructure is encapsulated with hBN on both sides to preserve the intrinsic properties of the layers \cite{dean2010boron}. From the top, (see Fig.\ref{1}a), the 2D layers are picked up in the following order: hBN-BLG-3R-MoS$_2$-hBN. Electrical leads were patterned by electron-beam lithography after reactive ion plasma etching in a hall bar geometry (see Fig.\ref{1}a). Thereafter, metal deposition of Cr/Au (5 nm/50 nm) was performed to form high-quality one-dimensional edge contacts onto bilayer graphene \cite{dean2010boron}. The top and bottom gate dielectrics are as follows: hBN as the top-gate dielectric (thickness $\approx$ 28 nm) and a global back-gate with hBN ($\approx$ 30 nm thickness) on SiO$_2$(285 nm), schematically shown in Fig.\ref{1}b. We have also fabricated graphene/2H-MoS$_{2}$ device in the same geometry as a control device. In both the cases, bilayer MoS$_2$ was chosen. The device details electrical characterization and optoelectronic response of all the devices studied are provided in Supporting Information. Photoconductance measurements were performed using a 532 nm continuous wave laser at a temperature of 100 K under ultrahigh vacuum conditions in a flow-type cryostat equipped with an optical window. The time-resolved photoresponse was recorded using a Data Acquisition Card (DAC) interfaced with lock-in amplifier. Power-dependent optoelectronic measurements are performed using a voltage-controlled attenuator in series with 532 nm laser.

\section{III. Results}
\subsection{Characterization of BLG-3R MoS$_{2}$ hybrid}
The 3R-MoS$_{2}$ was characterized using KPFM. A multilayered flake (as shown in an optical image Fig.\ref{1}c.) was chosen to map the surface potential at room temperature. The flake has a bilayer (BL) and trilayer (TriL) region, separated by a crack in between as indicated in the magnified image. Fig.\ref{1}c shows the potential map of the zoomed region - it reveals three regions with distinct contrast. The triL region - with a brighter contrast and the BL region, with two slightly different contrasts. This contrast difference in the BL corresponds to domains with opposite polarization orientations (either $AB$ domain or $BA$ domain). A line-cut across the map indicated by the arrow shows an interlayer potential difference ($\Delta V_{KPFM}$) of $\approx$ 70 mV, similar to previously observed values \cite{deb2022cumulative,weston2022interfacial}. To further confirm the presence of the 3R phase, photoluminescence (PL) spectroscopy was performed in 3R- and 2H-bilayer MoS$_2$ at room temperature revealing negatively charged trion ($A^-$), neural exciton $A$ and $B$ as shown in Fig.\ref{1}d. The PL spectra of 3R exhibit a redshift of the $A$ and $B$ excitonic transitions by 34 meV and 42 meV, respectively, in comparison to the 2H phase \cite{zhou2025identification}. This is attributed to reduction in quasi-particle bandgap outpacing the reduction in binding energy in 3R stacking \cite{he2014stacking}. In addition, the energy separation between the $A$ and $B$ excitons in the 3R-phase was found to be smaller by $\approx$ 8 meV relative to that of 2H-MoS$_2$ \cite{zhou2025identification}. The interlayer coupling between the two layers of a TMD can be well-characterized by low frequency Raman modes. Fig.\ref{1}e shows the shear mode ($S$) at a frequency of 23 cm$^{-1}$ and the layer-breathing mode ($LB$) at 40 cm$^{-1}$. The amplitude ratio of $LB$ to $S$ mode $>$ 1 in the 3R polytpe. The frequency difference between the two modes $LB$ and $S$ is much lower in 3R as compared to 2H (see Fig.\ref{1}e) \cite{zhou2025identification,sam2020probing,van2019stacking}. These characterization results collectively provide strong evidence for the existence of ferroelectricity in mechanically exfoliated 3R-MoS$_2$ flakes.

Low temperature electrical characterization of encapsulated BLG/3R-MoS$_{2}$ was carried out by measuring the longitudinal resistance ($R$) as a function of back-gate voltage (V$_{bg}$) at different top-gate voltages (V$_{tg}$) ranging from $-5$ V to $+5$ V (see Fig.\ref{1}f). It exhibits an increase in resistance at the Dirac point ($R_{Dirac}$) with increasing V$_{tg}$, which is a characteristic signature of bandgap opening in BLG under a displacement field $D$. The phase space (Fig.\ref{1}g) shows the locus of the Dirac point - it is determined purely by the geometric gate-channel capacitance with the slope =$ C_{tg}/C_{bg}$. This is distinct from the effect of an underlying polar moir\'{e} domain where the channel disintegrates into regions of different charge densities with increasing $D$ \cite{sett2024emergent}, or an abrupt jump in the resistance from the impact of polarization switching that changes the total carrier density in the channel \cite{wang2019multimechanism}, as has been observed before. The evolution of $R$ at charge neutrality point (CNP) with $D$ is shown in Fig.\ref{1}h. There is an asymmetric increase in $R_{Dirac}$ with the direction of $D$. This is indicative of the presence of polarization field ($E_p$) acting on the BLG. In addition to asymmetry, the minimum resistance at CNP (with $D$) is at $D$= - 0.058 V/nm (enlarged view shown in Fig.\ref{1}h) and it closely matches the interlayer potential of $\approx$ 55 meV across the bilayer 3R-MoS$_2$ \cite{wang2022interfacial}. However, the $R-V_{bg}$ curves exhibit negligible hysteresis (see SI, section 3), indicative of a minimal polarization modulation with $D$. We attribute this behavior to the absence of ferroelectric domain walls within the flake, which are otherwise known to serve as nucleation sites for polarization reversal with electric field spanning 0.09 V/nm to 0.21 V/nm \cite{liang2025resolving,yang2024non}. In single-domains of polarization in a flake, polarization switching is considerably challenging due to large nucleation energy \cite{wu2021sliding}. Thus, the underlying polarization influences the position of the Fermi level Energy ($E_F$) in BLG and possibly has a single-domain structure. This corroborates with our observation from KPFM where up to 15-20 $\mu$m lengths of the flake show uniform polarization.

\begin{figure*} 
\includegraphics[width=1\linewidth]{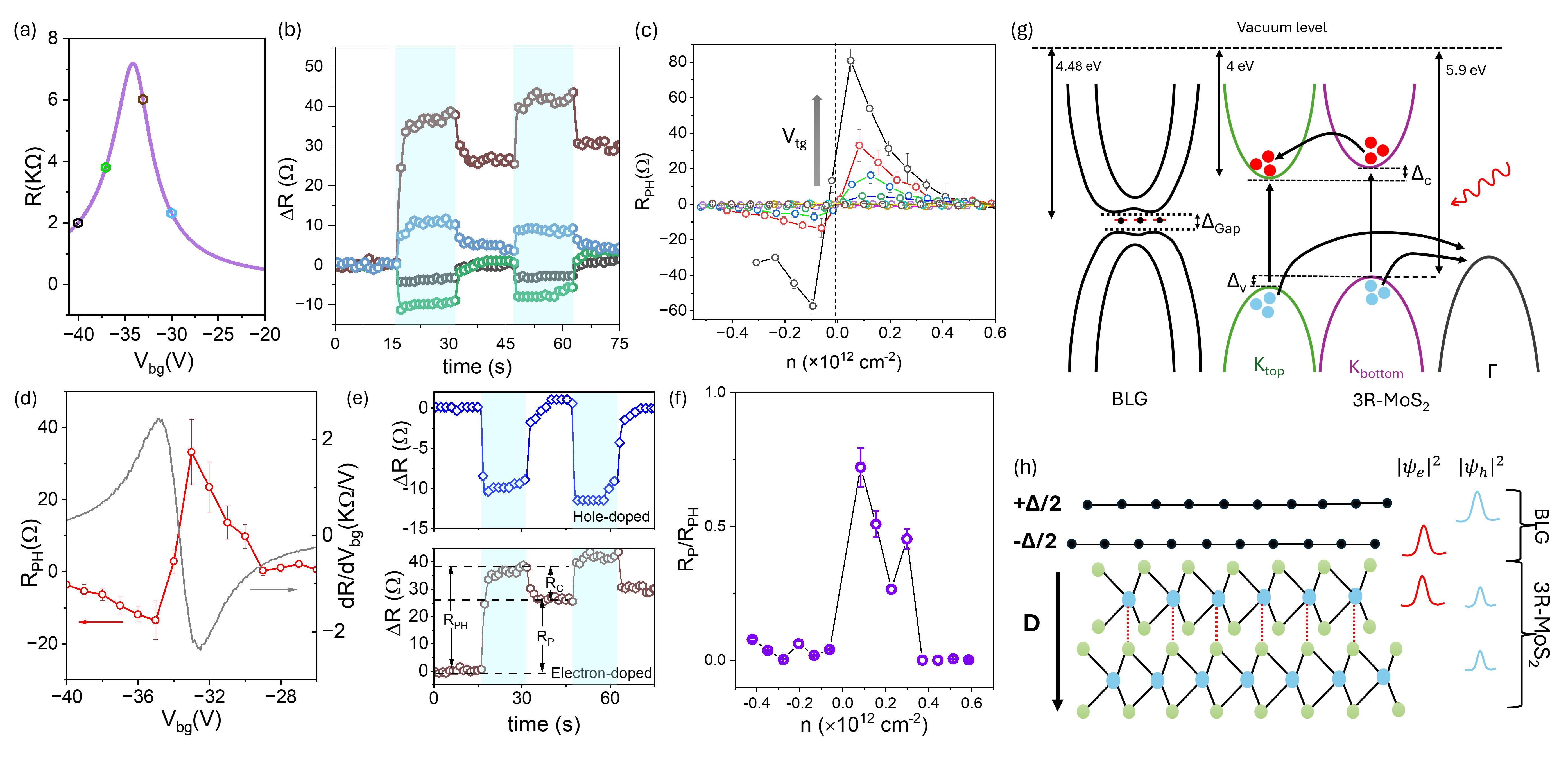} 
\caption{Photoresponse in BLG/3R-MoS$_{2}$ hybrid at $T$ = 100 K. (a) $R-V_{bg}$ curve at V$_{tg}$ = +4 V. (b) $\Delta R$ as a function of time in light "ON" and "OFF" ("ON" illustrated shaded region) at four different back-gate voltages ($V_{bg}$) highlighted by hollow circles in (a). (c) Photoresponse (R$_{PH}$) as a function of doping density ($n$) in BLG channel at different top-gate voltages ranging from -2 V to +5 V at an interval of +1 V. (d) R$_{PH}$ (red hollow circles) and dR/dV$_{bg}$ (solid grey line) as a function of $V_{bg}$ at V$_{tg}$ = +4 V. (e) $\Delta R$ as a function of time in hole-doped regime where V$_{bg}$-V$_{d}$= - 2.85 V (top panel) and electron- doped regime where V$_{bg}$-V$_{d}$= + 1.15 V (bottom panel). (f) $R_{P}/R_{PH}$ versus $n$ at V$_{tg}$ = +4 V. (g) Schematic of the hybrid at $D \ne 0$  illustrating the band gap opening in BLG and relaxation of electron and holes in 3R-MoS$_2$ under illumination. (h) Schematic of the layer polarization of electron and hole wavefunction in BLG and 3R-MoS$_2$ in the presence of $D$ pointing towards the substrate.} 
\label{2}
\end{figure*}

\subsection{Effect of Displacement field on the photoresponse in BLG/3R-MoS$_2$}
Having established the existence of polarization in 3R-MoS$_2$ and emergence of a electric field-induced bandgap opening in the BLG/3R-MoS$_2$ hybrid, we proceed to characterize the photoresponse of the heterostructures under an optical excitation of $\lambda$ = 532 nm. Fig.\ref{2}a presents a typical $R–V_{bg}$ curve acquired at $V_{tg}$ = 4 V. Fig.\ref{2}b shows the temporal photoresponse traces recorded under alternating light illumination ("ON") and dark ("OFF") conditions for four representative values of $V_{bg}$, corresponding to the operating points highlighted by circles. We systematically investigate the dependence of the photoresponse on $D$ across a range of $V_{tg}$ spanning from -2 V to +5 V (see Fig.\ref{2}c), corresponding to carrier doping densities ranging from $- 0.5\times 10^{12}$ cm$^{-2}$ to $+ 0.5\times 10^{12}$ cm$^{-2}$. The photoresponse shows a monotonic enhancement of magnitude with increasing $D$. We attribute this to the concomitant increase in $dR/dV_{bg}$ at higher values of $D$ (see SI, section 4) as consequence of the steeper resistance profile in BLG that amplifies the sensitivity of the channel to any photoinduced change in carrier density. This tunability of Responsivity ($R^*$, photocurrent generated per unit optical power) with $D$ provides an additional and independent experimental knob through which the photodetection performance of the device can be substantially enhanced \cite{roy2018number}.

\begin{figure*} 
\includegraphics[width=1\linewidth]{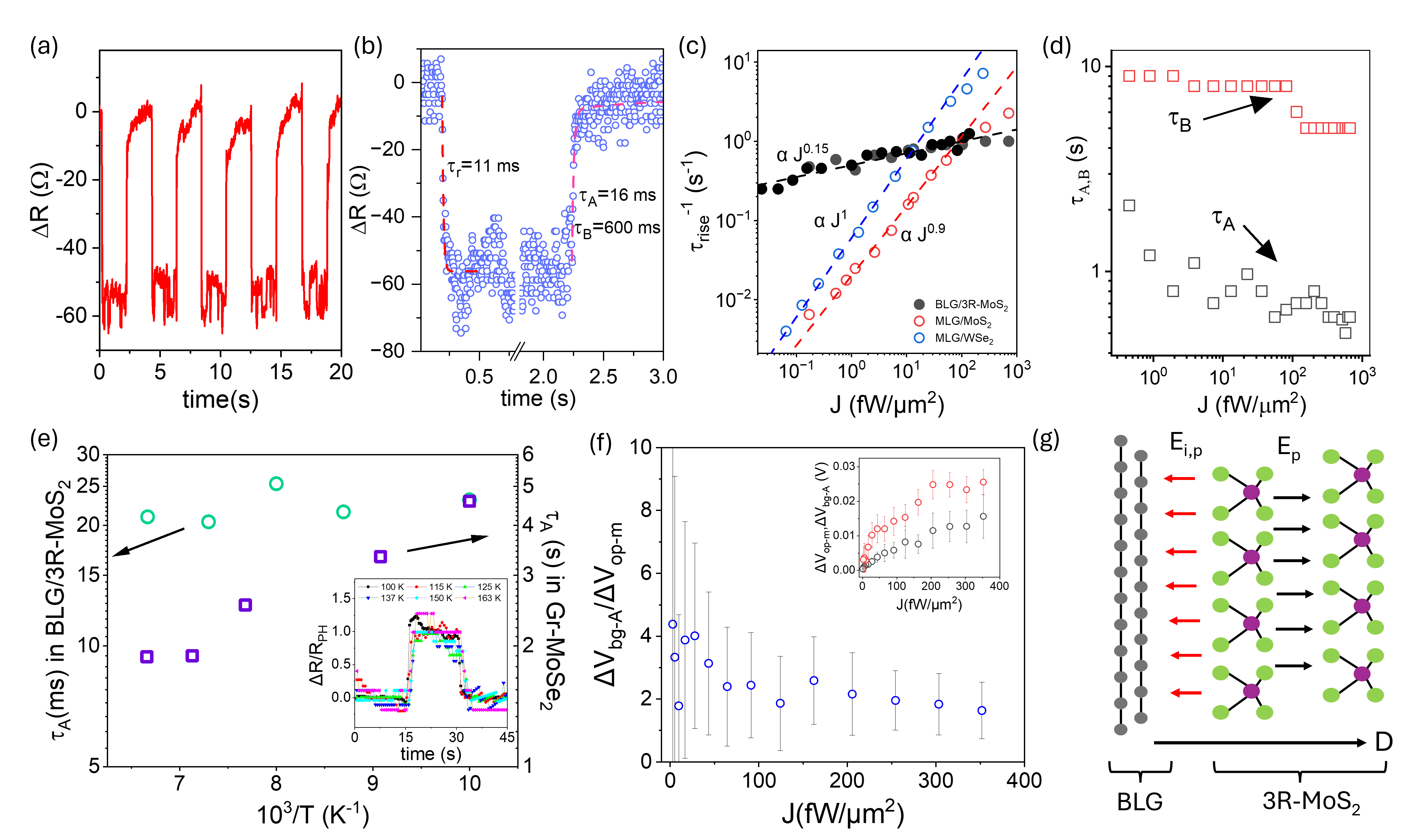}
\caption{Persistence-free photoresponse in hole-doped regime. (a) $\Delta$R as a function of time at $V_{bg}$ = - 43 V, $V_{tg}$ = +5 V for six optical pulses with a pulsing frequency of 0.5 Hz at 100 K. (b) A typical representation of $\Delta$R vs time for a single
optical pulse at $V_{bg}$ = -43 V, $V_{tg}$ = +5 V. The rise and fall region is fitted with a exponential and bi-exponential fit (red dash lines) respectively. (c) Power dependence of the inverse of rise time in BLG/3R-MoS$_2$ and other Gr/TMD hybrids \cite{kashid2020observation} (d)
Power dependence of fall time at $V_{bg}$ = -43 V, $V_{tg}$ = +5 V. The red and black hollow squares represent to two processes of fall timescales involved in recovery. (c) Recovery timescale $\tau_A$ and $\tau_B$ as a function of intensity $J$. (e) Temperature dependence of $\tau_A$ ranging from 100 K to 165 K for BLG/3R-MoS$_2$ (green solid circles) and Gr/MoSe$_2$ (purple hollow squares) \cite{parappurath2022interlayer}. The inset represents $\Delta R/\Delta R_{PH}$ as a function of time under one pulse of illumination at different
temperatures. (f) Ratio of gate potential change ($\Delta V$) due to illumination and from electrostatic considerations, plotted as a function of intensity. inset shows the individual gate potential changes, $\Delta V_{bg_A}$ and $\Delta V_{op_m}$ (g) Schematic of electric field directions in BLG/3R-MoS$_2$. $E_p$ is the polarization induced field at the interface of bilayer 3R-MoS$_2$, $E_{i,p}$ is the field at interface of BLG and 3R-MoS$_2 $ because of induced polarization, and $D$ is the displacement field.}
\label{3}
\end{figure*}

In Fig.\ref{2}d, the photoinduced resistance change ($R_{PH}$) and mathematically calculated $dR/dV_{bg}$ is shown as a function of $V_{bg}$. Notably, the sign of $R_{PH}$ suggests a net transfer of electrons from BLG to the TMD layer. This is contrary to the conventionally expected direction of photoinduced charge transfer in Gr/TMD based heterostructures \cite{roy2013graphene, parappurath2022interlayer}. Fig.\ref{2}e shows the photoresponse in electron-doped region and hole-doped regions separately. There is a striking asymmetry - we observe a persistence-free photoresponse on the hole-doped side, wherein the photoresistance returns to its pre-illumination baseline upon switching off the optical excitation. In contrast, the electron-doped regime shows a partially persistent ($R_P$) photoresponse, characterized by an incomplete recovery ($R_C$) following the termination of illumination (see Fig.\ref{2}e). This doping-asymmetric behavior is highlighted in Fig.\ref{2}f where we plot the ratio $R_{P}/R_{PH}$ as a function of doping density. We note that the degree of persistence in the photoresponse is governed by the position of $E_F$ in BLG. Persistence is confined to only a part of the electron-doped regime, (see SI, section 4), and is within the bandgap of BLG. In Gr–TMD hybrids, persistent photoresponse is overwhelmingly attributed to photogating caused by long-lived trapped carriers (in the TMD or at the graphene–TMD interface), producing a sustained electrostatic gating effect. However, persistence only near the CNP suggests that the localized density of states or trap distribution lies within the bandgap of BLG near conduction band (as schematically demonstrated in Fig.\ref{2}g). The opposite sign of $R_{PH}$ and $dR/dV_{bg}$ can be explained in terms of charge-reorganization in 3R-MoS$_2$. A higher probability density of holes $|\psi|_h^2 $ in the top MoS$_2$ layer (as shown in Fig.\ref{2}h), and occupation of holes in the top graphene layer due to $D$-induced layer polarization in BLG leads to this effect as discussed later in section IV.

\subsection{Persistence-free photoresponse in the hole-doped regime}
 Fig.\ref{3}a shows a representative temporal photoresponse trace recorded over multiple alternating illumination ("ON") and dark ("OFF") cycles of duration 2 s each, acquired in the persistence free hole-doped regime ($V_{bg}$ = -43 V, $V_{tg}$ = 5 V, $D$ = 0.44 V/nm). The response time, $\tau$ were extracted by fitting the raw temporal data (see Fig.\ref{3}b). The rise time ($\tau_{rise}$) is 9.5$\pm$2 ms (instrument-limited). We systematically investigated the persistence-free photoresponse as a function of incident optical power density ($J$), ranging from 0.1 fW/$\mu$m$^2$ to 800 fW/$\mu$m$^2$. Fig.\ref{3}c shows inverse of extracted rise time as a function of $J$. There is a remarkably weak dependence while Gr/TMD hybrids (shown by filled circle) show an almost linear dependence \cite{parappurath2022interlayer}, suggesting that the underlying relaxation mechanism in BLG/3R-MoS$_2$ is not governed by the conventional trap-mediated recombination processes \cite{ahmed2020interplay}.

The decay dynamics exhibit a complex behavior well-described by a bi-exponential fit as given below:
 \begin{equation}
 \Delta R = R_A (1-e^{(t-t_0)/\tau _A}) + R_B (1-e^{(t-t_0)/\tau _B}),
\end{equation} 
 where $R_A$, $R_B$ are the contribution of two different relaxation pathways and $\tau_A$, $\tau_B$ correspond to timescales of two relaxation pathways. The fit (dotted red line in Fig.\ref{3}b)  comprises a fast decay component which contributes to $\approx$ 90$\%$ of photoresponse ($\tau_A$) with a timescale of 23$\pm$5.8 ms and a slower component ($\tau_B$) with a timescale of 600$\pm$37.8 ms contributing around 10$\%$ to photoresponse. Both $\tau_A$ and $\tau_B$ vary negligibly over four orders of optical illumination (Fig.\ref{3}d) (see SI for data collection procedure). The relative contributions of fast and slow photoresponse components ($R_A$ and $R_B$) to the total photoresponse signal are found to scale sub-linearly with optical power. The fast component, $R_A$ exhibiting a marginally steeper power dependence compared to the slow component $R_B$ (see SI, section 7). This behavior is characteristic of systems in which both shallow and deep trap states are present simultaneously, wherein recombination centres and shallow traps govern the fast relaxation channel and deep traps mediate the slower, power-independent recombination pathways \cite{xu2025role}.

To understand the underlying charge transfer mechanism, we investigate the temperature dependence of $\tau_A$ as shown in Fig.\ref{3}e. Strikingly, $\tau_A$ exhibits no discernible dependence on temperature (green solid circles), which is in contrast to the thermally activated behavior reported previously for Gr/MoSe$_2$ hybrids where the fall times decreases with increasing temperature (blue squares in Fig.\ref{3}e) \cite{parappurath2022interlayer}. The absence of any appreciable thermal activation in $\tau_A$ points towards an underlying mechanism that is fundamentally independent of thermal energy, possibly related to charge-reorganization in 3R-MoS$_2$ (as discussed in Section IV).

\begin{figure*} 
\includegraphics[width=1\linewidth]{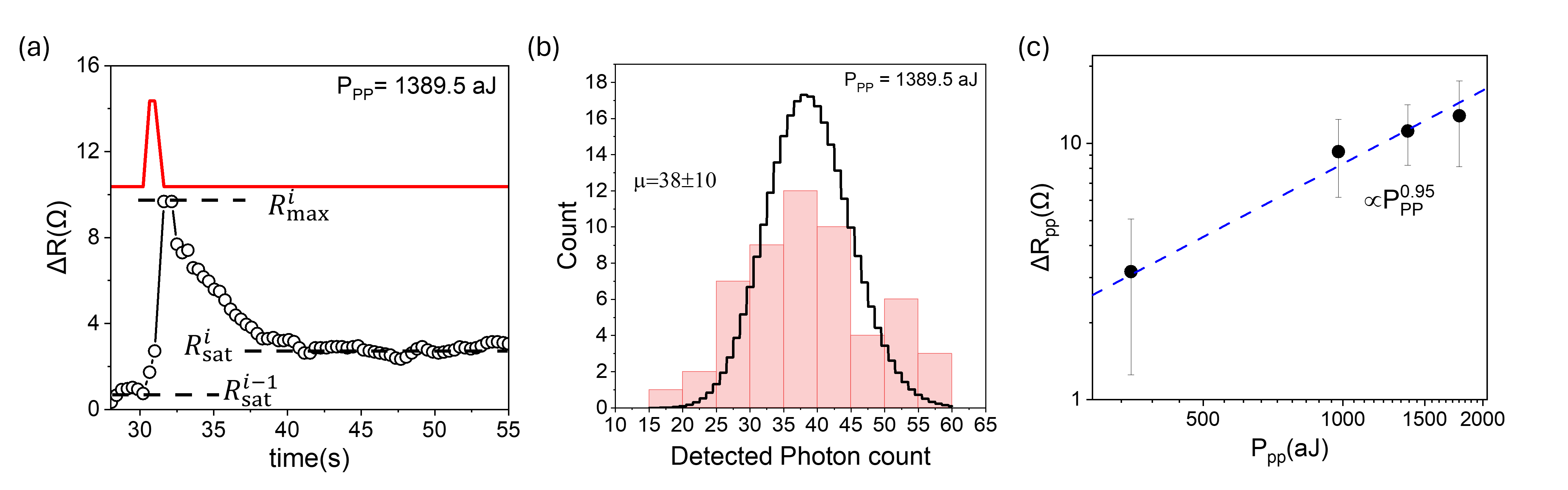} 
\caption{Ultra-low light detection with BLG/3R-MoS$_2$ hybrid. (a) Resistance as a function of time after a single pulse of light at 1389.5 aJ.  (b) Histogram of normalized photon numbers obtained from multiple sets of pulsing experiments as shown in (a).(c) Average change in resistance ($\Delta R_{PP}$ as a function of power per pulse ($P_{PP}$) ranging from 350 aJ to 2000 aJ. The blue dotted line is the power law fit to the data.} 
\label{4}
\end{figure*}

\subsection{Ultra-low light detection in the persistent regime}
 The systematic enhancement of photoresponse with increasing $D$ in BLG/3R-MoS$_{2}$ heterostructure provides an additional parameter for optimizing sensitivity of the photodetector. Fig.\ref{4}a show the photoresistance (after processing using the moving-point average) acquired for an illumination pulses of temporal width 1 s (at an interval of 30 s between successive pulses) in the electron-doped regime (at $V_{tg}$ = 6.5 V and $V_{bg}$ = -53 V). The resistance first rises to a peak value R$^{i}_{max}$ and then relaxes to a steady value R$^{i}_{sat}$, where $i$ denotes the $i^{th}$ pulse (Fig.\ref{4}a). This rise in resistance coincides with the rising edge of the pulse. The change in resistance per pulse is computed as, $\Delta R^{i}_{pp} = R^{i}_{max} - R^{i-1}_{sat}$. Then $\Delta R^{i}_{pp}$, due to illumination is converted to number of photons ($N$) detected by the following relation,
\begin{equation}
 N = \Delta R^{i}_{pp}/ \Delta R_{1e},
\label{eq:1}
\end{equation} 
\begin{equation}
 \Delta R_{1e} = dR/dV_{bg} \times \Delta V^{1e}_{bg};
 \label{eq:2}
\end{equation}
where $\Delta V^{1e}_{bg} = e/C_{bg}$, is the change in gating voltage corresponding to one charge carrier.

For the photon counting experiments, the sample was irradiated with six pulses of identical ultra-low powers (down to 350 aJ) in a cycle. After each cycle, the device was brought to its initial state by sweeping the gate voltage. Fig.\ref{4}b shows the detected photon events from these measurements accumulated into a histogram that was fit to a Poisson distribution and shows the number of detected photons per pulse. From the fits, the smallest mean photon count ($\mu$) that can be reliably detected is approximately 31 photons (see SI, section 6), with an uncertainty of $\pm$ 10 photons.
Using these values, the internal quantum efficiency (IQE), defined as the ratio of the number of detected photons to the number of absorbed photons, is found to be  10 $\%$ for incident power of 1389.5 aJ. In Fig.\ref{4}c $\Delta R_{pp}$ is shown as a function of power per pulse ($P_{PP}$) and it indicates a linear regime of photoresponse. This is supported by near-constant IQE over this range of incident powers indicates that the device operates within a linear detection regime (see SI, section 6). This device simultaneously achieves high gain and relatively fast response, making it especially well‑suited for repetitive low‑intensity optical signal detection due to photoinduced polarization modulation in 3R-MoS$_2$.

 Gain–bandwidth product (GBP) is figure of merit ($G = R^{*} hc/(e \lambda \eta)$, $\eta$ is the quantum efficiency, bandwidth (BW) is limited by the fall time and approximated as $0.35/\tau$. Our device exhibits one of the highest reported GBW, $\approx 5 \times 10^8$ Hz, with a photogain of $3.1 \times 10^7$ and a fall‑time‑limited bandwidth of 15.9 Hz (instrument limited). See SI section 9 for a comparative plot.

\section{IV. Discussions}
The kinetics of interfacial charge transfer at BLG/3R-MoS$_{2}$ interface is markedly different compared to Gr/TMD hybrids - the observed anomalous effects are - (1) Asymmetry in photoresponse in the electron and hole doped regime  (2) Persistence-free photoresponse at low temperatures (3) Relatively faster recovery timescales of 10's of ms as compared to 100's of seconds in Gr/TMD hybrids at low temperature and (5) temperature independent fall timescales. We attribute these factors to the influence of the polarization-induced electric field of 3R-MoS$_2$ on the charge-transfer dynamics at the interface as we discuss below.

Contrary to inversion-symmetric TMD layers, 3R-stacked bilayer has a type-II band alignment at $K$ point owing to the interlayer potential across two layers with conduction band minima (CBM) localized in the top MoS$_{2}$ layer and the valence band maxima (VBM) residing at $\Gamma$ point (illustrated in Fig.\ref{2}g) \cite{liang2022optically} naturally resulting in layer selective confinement of electron and hole wavefunctions.
Under illumination, photoinduced doping lowers the polarization in 3R-MoS$_2$ \cite{deb2022cumulative, gao2024large} and thereby reduces the interfacial field ($E_{i,P}$) at the BLG/MoS$_2$ junction. As a result, the effective $D$ increases and shifts the $E_F$ downward in BLG channel. To substantiate this argument, we compare the change in gate potential experienced by BLG at $V_{bg}$ = -57 V and $V_{tg}$= +6.5 V under illumination ($\Delta V_{op-m}$) arising from photoinduced redistribution of charges in 3R-MoS$_2$ (see inset of Fig.\ref{3}f, black hollow circles) with the gate potential change extracted from $R_{A}$ ($\Delta V_{bg-A}= R_{A}/ (dR/dV_{BG})$; more details on these calculations are given in SI, section 8). The $\Delta V_{bg-A}$ is of similar order of magnitude as $\Delta V_{op-m}$ even though there are uncertainties in absorption coefficient magnitude and variation in optical transmission through top metal gate electrode that we have used for the calculations. This reduction in net gate potential, manifests as a negative photoresistance in the hole-doped regime and a positive photoresistance in the electron-doped regime. Thus, the effective $D$ experienced by BLG channel is the sum of the $D$ and $E_{i,p}$ at the interface, as illustrated in Fig.~\ref{3}g. We ensure device operation in a regime where $V_{tg}$ $>$ 0 and $V_{bg}$ is $<$ 0, such that $D$ is consistently directed downward, away from the BLG, towards the substrate. Thus, the observed sign reversal of the photoresponse, coupled with the photoinduced reduction in polarization, identifies 3R-MoS$_2$ as the origin of the observed behavior. 

The asymmetry in persistence across hole-doped and electron-doped regime is a consequence of layer polarization in BLG. When $D$ points downward, it generates an interlayer potential difference $\Delta$, such that the top layer is at a higher potential of $+\Delta/2$ and vice versa for bottom layer. The VBM gets localized in the top layer leading to a higher probability density of holes ($|\psi|_h^2$) (see Fig \ref{2}h). In the hole-doped regime, electronic transport occurs predominantly through the top graphene layer and since the charge carriers are farther away from the interface, charge transfer between BLG and 3R-MoS$_2$ is suppressed. Thus, the observed photoresponse arises primarily from polarization-modulation and is persistence free. In the electron-doped regime, transport is mainly through the bottom layer of BLG, which is closer to the BLG/3R-MoS$_2$ interface, facilitating charge transfer and leading to partial persistence.  The observed partial persistence is restricted to an energy window on the electron-doped regime with $E_F$ within the bandgap, promoting trapping of photogenerated electrons and giving rise to a persistent photoresponse. The maximum $R^*$ obtained in the partial persistence mode is 1.6$\times$10$^7$ A/W. Moreover, there is a lower density of localized trap states at the Gr/TMD interface in bilayer configuration due to preferential vacancy distribution of interlayer interface of the bilayer configuration \cite{tan2020stability,zhou2017atomic,wan2026defect}. Thus, these mechanisms work collectively resulting in persistent free photoresponse.

Previous studies on trap-mediated photoresponse reported that the inverse rise time scales with incident power density shows thermally activated behaviour \cite{parappurath2022interlayer,kashid2020observation}. In contrast, we observe only a weak dependence on power (see Fig\ref{3}c), supporting our claim that the timescales of $\approx$ few tens of ms in our device originate from a mechanism largely independent of both illumination power and temperature—specifically, polarization modulation driven by photoinduced doping. At very low powers ($<$ 0.5 fW/$\mu$m$^2$), the rise time is about 2–3 s, indicating that in this regime trap-mediated charge transfer dominates; however, at powers ($>$ 1 fW/$\mu$m$^2$), polarization modulation becomes the primary contributor to the photoresponse.

\section{V. Conclusions}
In summary, the uniform ferroelectric polarization inherent to the 3R-MoS$_2$ layer plays a central and versatile role in reconfiguring the interfacial potential landscape: it simultaneously functions as a ferroelectric gate that can be optically tuned to deliver a persistence-free optoelectronic response, while also screening residual trap potentials at the Gr/TMD interface. Together, these effects suppress the trap-mediated persistence pathways that typically govern non-ferroelectric TMD-based heterostructures. By breaking layer inversion symmetry in MoS$_2$ and thereby inducing a spontaneous out-of-plane polarization, we obtain not only a persistence-free response but also accelerated response times $\approx$ 11 ms (instrument-limited), overcoming the trade-off between high responsivity and slow photodetection in charge-transfer-driven conventional optoelectronic photodetectors. The dependence of the photoresponse on the position of $E_F$ thus provides a tunable knob to operate the photodetector in two distinct modes: a repeatable photodetection mode, where $E_F$ is set in hole-doped regime with purely non-persistent response, and an optical non-volatile memory mode, where $E_F$ is tuned into electron doped regime to realize a non-volatile photoresponse. Our findings underscore the promise of ferroelectric layers for realizing multifunctional optoelectronic platforms that couple high-speed performance with non-volatile optical memory, thereby creating new possibilities for optical communication and computing technologies.

\section*{Author Contributions}
N.K.G and S.S conceived and designed the experiments. N.K.G performed the experiments, analyzed the data and wrote the manuscript. S.S partially performed experiments, edited and revised the manuscript. S.K partially performed data acquisition. K.W. and T.T, provided partial support in material synthesis. A.G. supervised the overall project.

\section*{Competing Interests}
The authors declare no competing interests.

\section*{Acknowledgments}
The authors acknowledge help from Dr. Krishnendu Dandapat, IISc Bangalore and Ansh Gupta, IIT Bhubaneshwar. The authors also acknowledge NNFC facilities at CeNSE, IISc Bangalore. A.G. acknowledges support from funded project under DST Nanomission, India. N.K.G acknowledges funding from SP/AMIO-25-0001. K.W. and T.T. acknowledge JSPS KAKENHI (21H05233, 23H02052), CREST (JPMJCR24A5), JST, and WPI-MEXT.

\section*{Data Availability Statement}
The data that support the findings of this study is available from the corresponding author upon reasonable request.

\bibliography{References.bib}

@article{roy2013graphene,
  title={Graphene--MoS2 hybrid structures for multifunctional photoresponsive memory devices},
  author={Roy, Kallol and Padmanabhan, Medini and Goswami, Srijit and Sai, T Phanindra and Ramalingam, Gopalakrishnan and Raghavan, Srinivasan and Ghosh, Arindam},
  journal={Nature nanotechnology},
  volume={8},
  number={11},
  pages={826--830},
  year={2013},
  publisher={Nature Publishing Group UK London}
}

@article{parappurath2022interlayer,
  title={Interlayer charge transfer and photodetection efficiency of graphene--transition-metal-dichalcogenide heterostructures},
  author={Parappurath, Aparna and Mitra, Sreemanta and Singh, Gagandeep and Gill, Navkiranjot Kaur and Ahmed, Tanweer and Sai, T Phanindra and Watanabe, Kenji and Taniguchi, Takashi and Ghosh, Arindam},
  journal={Physical Review Applied},
  volume={17},
  number={6},
  pages={064062},
  year={2022},
  publisher={APS}
}

@article{kashid2020observation,
  title={Observation of inter-layer charge transmission resonance at optically excited graphene--TMDC interfaces},
  author={Kashid, Ranjit and Mishra, Jayanta Kumar and Pradhan, Avradip and Ahmed, Tanweer and Kakkar, Saloni and Mundada, Pranav and Deshpande, Preeti and Roy, Kallol and Ghosh, Ambarish and Ghosh, Arindam},
  journal={APL Materials},
  volume={8},
  number={9},
  year={2020},
  publisher={AIP Publishing}
}

@article{gill2025moire,
  title={Moir{\'e} Ferroelectricity-Enhanced Optoelectronic Response in an all-2D van der Waals Hybrid},
  author={Gill, Navkiranjot Kaur and Sett, Shaili and Debnath, Rahul and Singha, Arup and Watanabe, Kenji and Taniguchi, Takashi and Ghosh, Arindam},
  journal={Small},
  volume={21},
  number={37},
  pages={e05797},
  year={2025},
  publisher={Wiley Online Library}
}

@article{ahmed2020interplay,
  title={Interplay of charge transfer and disorder in optoelectronic response in Graphene/hBN/MoS2 van der Waals heterostructures},
  author={Ahmed, Tanweer and Roy, Kallol and Kakkar, Saloni and Pradhan, Avradip and Ghosh, Arindam},
  journal={2D Materials},
  volume={7},
  number={2},
  pages={025043},
  year={2020},
  publisher={IOP Publishing}
}

@article{sett2024emergent,
  title={Emergent inhomogeneity and nonlocality in a graphene field-effect transistor on a near-parallel Moir{\'e} superlattice of transition metal dichalcogenides},
  author={Sett, Shaili and Debnath, Rahul and Singha, Arup and Mandal, Shinjan and Jyothsna, KM and Bhakar, Monika and Watanabe, Kenji and Taniguchi, Takashi and Raghunathan, Varun and Sheet, Goutam and others},
  journal={Nano Letters},
  volume={24},
  number={30},
  pages={9245--9252},
  year={2024},
  publisher={ACS Publications}
}

@article{wang2022interfacial,
  title={Interfacial ferroelectricity in rhombohedral-stacked bilayer transition metal dichalcogenides},
  author={Wang, Xirui and Yasuda, Kenji and Zhang, Yang and Liu, Song and Watanabe, Kenji and Taniguchi, Takashi and Hone, James and Fu, Liang and Jarillo-Herrero, Pablo},
  journal={Nature nanotechnology},
  volume={17},
  number={4},
  pages={367--371},
  year={2022},
  publisher={Nature Publishing Group UK London}
}

@article{weston2022interfacial,
  title={Interfacial ferroelectricity in marginally twisted 2D semiconductors},
  author={Weston, Astrid and Castanon, Eli G and Enaldiev, Vladimir and Ferreira, Fabio and Bhattacharjee, Shubhadeep and Xu, Shuigang and Corte-Le{\'o}n, H{\'e}ctor and Wu, Zefei and Clark, Nicholas and Summerfield, Alex and others},
  journal={Nature nanotechnology},
  volume={17},
  number={4},
  pages={390--395},
  year={2022},
  publisher={Nature Publishing Group UK London}
}

@article{deb2022cumulative,
  title={Cumulative polarization in conductive interfacial ferroelectrics},
  author={Deb, Swarup and Cao, Wei and Raab, Noam and Watanabe, Kenji and Taniguchi, Takashi and Goldstein, Moshe and Kronik, Leeor and Urbakh, Michael and Hod, Oded and Ben Shalom, Moshe},
  journal={Nature},
  volume={612},
  number={7940},
  pages={465--469},
  year={2022},
  publisher={Nature Publishing Group UK London}
}

@article{liang2022optically,
  title={Optically probing the asymmetric interlayer coupling in rhombohedral-stacked MoS 2 bilayer},
  author={Liang, Jing and Yang, Dongyang and Wu, Jingda and Dadap, Jerry I and Watanabe, Kenji and Taniguchi, Takashi and Ye, Ziliang},
  journal={Physical Review X},
  volume={12},
  number={4},
  pages={041005},
  year={2022},
  publisher={APS}
}

@article{scuri2020electrically,
  title={Electrically tunable valley dynamics in twisted WSe 2/WSe 2 bilayers},
  author={Scuri, Giovanni and Andersen, Trond I and Zhou, You and Wild, Dominik S and Sung, Jiho and Gelly, Ryan J and B{\'e}rub{\'e}, Damien and Heo, Hoseok and Shao, Linbo and Joe, Andrew Y and others},
  journal={Physical review letters},
  volume={124},
  number={21},
  pages={217403},
  year={2020},
  publisher={APS}
}

@article{choi2021twist,
  title={Twist angle-dependent interlayer exciton lifetimes in van der Waals heterostructures},
  author={Choi, Junho and Florian, Matthias and Steinhoff, Alexander and Erben, Daniel and Tran, Kha and Kim, Dong Seob and Sun, Liuyang and Quan, Jiamin and Claassen, Robert and Majumder, Somak and others},
  journal={Physical Review Letters},
  volume={126},
  number={4},
  pages={047401},
  year={2021},
  publisher={APS}
}

@article{wan2026defect,
  title={Defect-mediated carrier trapping and nonradiative recombination in two-dimensional sliding ferroelectrics},
  author={Wan, Honghao and Yu, Jianxin and Yang, Kun and Zhu, Yuanhao and Li, Jia-Wen and Fu, Huixia and Shi, Xinghua and Zhang, Jin},
  journal={The Journal of Chemical Physics},
  volume={164},
  number={9},
  year={2026},
  publisher={AIP Publishing}
}

@article{yang2022spontaneous,
  title={Spontaneous-polarization-induced photovoltaic effect in rhombohedrally stacked MoS2},
  author={Yang, Dongyang and Wu, Jingda and Zhou, Benjamin T and Liang, Jing and Ideue, Toshiya and Siu, Teri and Awan, Kashif Masud and Watanabe, Kenji and Taniguchi, Takashi and Iwasa, Yoshihiro and others},
  journal={Nature Photonics},
  volume={16},
  number={6},
  pages={469--474},
  year={2022},
  publisher={Nature Publishing Group UK London}
}

@article{wang2017interlayer,
  title={Interlayer coupling in commensurate and incommensurate bilayer structures of transition-metal dichalcogenides},
  author={Wang, Yong and Wang, Zhan and Yao, Wang and Liu, Gui-Bin and Yu, Hongyi},
  journal={Physical Review B},
  volume={95},
  number={11},
  pages={115429},
  year={2017},
  publisher={APS}
}

@article{ko2023operando,
  title={Operando electron microscopy investigation of polar domain dynamics in twisted van der Waals homobilayers},
  author={Ko, Kahyun and Yuk, Ayoung and Engelke, Rebecca and Carr, Stephen and Kim, Junhyung and Park, Daesung and Heo, Hoseok and Kim, Hyun-Mi and Kim, Seul-Gi and Kim, Hyeongkeun and others},
  journal={Nature Materials},
  volume={22},
  number={8},
  pages={992--998},
  year={2023},
  publisher={Nature Publishing Group UK London}
}

@article{li2017binary,
  title={Binary compound bilayer and multilayer with vertical polarizations: two-dimensional ferroelectrics, multiferroics, and nanogenerators},
  author={Li, Lei and Wu, Menghao},
  journal={ACS nano},
  volume={11},
  number={6},
  pages={6382--6388},
  year={2017},
  publisher={ACS Publications}
}

@article{he2014stacking,
  title={Stacking effects on the electronic and optical properties of bilayer transition metal dichalcogenides MoS 2, MoSe 2, WS 2, and WSe 2},
  author={He, Jiangang and Hummer, Kerstin and Franchini, Cesare},
  journal={Physical Review B},
  volume={89},
  number={7},
  pages={075409},
  year={2014},
  publisher={APS}
}

@article{zhou2025identification,
  title={Identification of polytypism and their dislocations in bilayer MoS2 using correlative transmission electron microscopy and Raman spectroscopy},
  author={Zhou, Xin and Dierke, Tobias and Wu, Mingjian and You, Shengbo and G{\"o}tz, Klaus and Unruh, Tobias and Pelz, Philipp and Will, Johannes and Maultzsch, Janina and Spiecker, Erdmann},
  journal={npj 2D Materials and Applications},
  volume={9},
  number={1},
  pages={58},
  year={2025},
  publisher={Nature Publishing Group UK London}
}

@article{sam2020probing,
  title={Probing stacking configurations in a few layered MoS2 by low frequency Raman spectroscopy},
  author={Sam, Rhea Thankam and Umakoshi, Takayuki and Verma, Prabhat},
  journal={Scientific Reports},
  volume={10},
  number={1},
  pages={21227},
  year={2020},
  publisher={Nature Publishing Group UK London}
}

@article{van2019stacking,
  title={Stacking-dependent interlayer phonons in 3R and 2H MoS2},
  author={Van Baren, Jeremiah and Ye, Gaihua and Yan, Jia-An and Ye, Zhipeng and Rezaie, Pouyan and Yu, Peng and Liu, Zheng and He, Rui and Lui, Chun Hung},
  journal={2D Materials},
  volume={6},
  number={2},
  pages={025022},
  year={2019},
  publisher={IOP Publishing}
}

@article{liang2025resolving,
  title={Resolving polarization switching pathways of sliding ferroelectricity in trilayer 3R-MoS2},
  author={Liang, Jing and Yang, Dongyang and Wu, Jingda and Xiao, Yunhuan and Watanabe, Kenji and Taniguchi, Takashi and Dadap, Jerry I and Ye, Ziliang},
  journal={Nature Nanotechnology},
  volume={20},
  number={4},
  pages={500--506},
  year={2025},
  publisher={Nature Publishing Group UK London}
}

@article{yang2024non,
  title={Non-volatile electrical polarization switching via domain wall release in 3R-MoS2 bilayer},
  author={Yang, Dongyang and Liang, Jing and Wu, Jingda and Xiao, Yunhuan and Dadap, Jerry I and Watanabe, Kenji and Taniguchi, Takashi and Ye, Ziliang},
  journal={Nature Communications},
  volume={15},
  number={1},
  pages={1389},
  year={2024},
  publisher={Nature Publishing Group UK London}
}

@article{wu2021sliding,
  title={Sliding ferroelectricity in 2D van der Waals materials: Related physics and future opportunities},
  author={Wu, Menghao and Li, Ju},
  journal={Proceedings of the National Academy of Sciences},
  volume={118},
  number={50},
  pages={e2115703118},
  year={2021},
  publisher={National Academy of Sciences}
}

@article{roy2018number,
  title={Number-Resolved Single-Photon Detection with Ultralow Noise van der Waals Hybrid},
  author={Roy, Kallol and Ahmed, Tanweer and Dubey, Harshit and Sai, T Phanindra and Kashid, Ranjit and Maliakal, Shruti and Hsieh, Kimberly and Shamim, Saquib and Ghosh, Arindam},
  journal={Advanced Materials},
  volume={30},
  number={2},
  pages={1704412},
  year={2018},
  publisher={Wiley Online Library}
}

@article{mitra2020graphene,
  title={Graphene-WS 2 van der Waals hybrid heterostructure for photodetector and memory device applications},
  author={Mitra, Sreemanta and Kakkar, Saloni and Ahmed, Tanweer and Ghosh, Arindam},
  journal={Physical Review Applied},
  volume={14},
  number={6},
  pages={064029},
  year={2020},
  publisher={APS}
}

@article{wang2013one,
  title={One-dimensional electrical contact to a two-dimensional material},
  author={Wang, Lei and Meric, I and Huang, PY and Gao, Q and Gao, Y and Tran, H and Taniguchi, T and Watanabe, Kenji and Campos, LM and Muller, DA and others},
  journal={Science},
  volume={342},
  number={6158},
  pages={614--617},
  year={2013},
  publisher={American Association for the Advancement of Science}
}

@article{dean2010boron,
  title={Boron nitride substrates for high-quality graphene electronics},
  author={Dean, Cory R and Young, Andrea F and Meric, Inanc and Lee, Chris and Wang, Lei and Sorgenfrei, Sebastian and Watanabe, Kenji and Taniguchi, Takashi and Kim, Phillip and Shepard, Kenneth L and others},
  journal={Nature nanotechnology},
  volume={5},
  number={10},
  pages={722--726},
  year={2010},
  publisher={Nature Publishing Group}
}

@article{novoselov2004electric,
  title={Electric field effect in atomically thin carbon films},
  author={Novoselov, Kostya S and Geim, Andre K and Morozov, Sergei V and Jiang, De-eng and Zhang, Yanshui and Dubonos, Sergey V and Grigorieva, Irina V and Firsov, Alexandr A},
  journal={science},
  volume={306},
  number={5696},
  pages={666--669},
  year={2004},
  publisher={American Association for the Advancement of Science}
}

@article{tan2020stability,
  title={Stability of charged sulfur vacancies in 2D and bulk MoS 2 from plane-wave density functional theory with electrostatic corrections},
  author={Tan, Anne Marie Z and Freysoldt, Christoph and Hennig, Richard G},
  journal={Physical Review Materials},
  volume={4},
  number={6},
  pages={064004},
  year={2020},
  publisher={APS}
}

@article{zhou2017atomic,
  title={Atomic structure and dynamics of defects in 2D MoS2 bilayers},
  author={Zhou, Si and Wang, Shanshan and Li, Huashan and Xu, Wenshuo and Gong, Chuncheng and Grossman, Jeffrey C and Warner, Jamie H},
  journal={ACS omega},
  volume={2},
  number={7},
  pages={3315--3324},
  year={2017},
  publisher={ACS Publications}
}

@article{wang2019multimechanism,
  title={Multimechanism synergistic photodetectors with ultrabroad spectrum response from 375 nm to 10 $\mu$m},
  author={Wang, Xudong and Shen, Hong and Chen, Yan and Wu, Guangjian and Wang, Peng and Xia, Hui and Lin, Tie and Zhou, Peng and Hu, Weida and Meng, Xiangjian and others},
  journal={Advanced science},
  volume={6},
  number={15},
  pages={1901050},
  year={2019},
  publisher={Wiley Online Library}
}

@article{bernardi2013extraordinary,
  title={Extraordinary sunlight absorption and one nanometer thick photovoltaics using two-dimensional monolayer materials},
  author={Bernardi, Marco and Palummo, Maurizia and Grossman, Jeffrey C},
  journal={Nano letters},
  volume={13},
  number={8},
  pages={3664--3670},
  year={2013},
  publisher={ACS Publications}
}

@article{wu2022ultrafast,
  title={Ultrafast response of spontaneous photovoltaic effect in 3R-MoS2--based heterostructures},
  author={Wu, Jingda and Yang, Dongyang and Liang, Jing and Werner, Max and Ostroumov, Evgeny and Xiao, Yunhuan and Watanabe, Kenji and Taniguchi, Takashi and Dadap, Jerry I and Jones, David and others},
  journal={Science Advances},
  volume={8},
  number={50},
  pages={eade3759},
  year={2022},
  publisher={American Association for the Advancement of Science}
}

@article{xu2025role,
  title={The role of trap states in MoS 2-based photodetectors},
  author={Xu, Yuhang and Wang, Yuxin and Zhang, Chunchi and Wu, Haijuan and Tan, Chao and Hu, Guohua and Wang, Zegao},
  journal={Nanoscale},
  volume={17},
  number={15},
  pages={9245--9252},
  year={2025},
  publisher={Royal Society of Chemistry}
}

@article{gao2024large,
  title={Large photoinduced tuning of ferroelectricity in sliding ferroelectrics},
  author={Gao, Lingyuan and Bellaiche, Laurent},
  journal={Physical Review Letters},
  volume={133},
  number={19},
  pages={196801},
  year={2024},
  publisher={APS}
}

@article{sett2025van,
  title={van der Waals Hybrids for Ferroelectric Device Application},
  author={Sett, Shaili and Paul, Tathagata and Ghosh, Arindam},
  journal={Annual Review of Materials Research},
  volume={55},
  year={2025},
  publisher={Annual Reviews}
}
\newpage
\end{document}